\documentclass[preprintnumbers,amsmath,amssymb,onecolumn,cite]{revtex4}
\usepackage[colorlinks=true,citecolor=red,filecolor=green,linkcolor=blue,pdfnewwindow=true]{hyperref}
\usepackage{color}
\usepackage{graphicx}
\usepackage{makeidx}
\usepackage{bm}
\usepackage[title]{appendix}
\usepackage{tabu}
\usepackage{float}
\usepackage[version=4]{mhchem}
\usepackage[ruled,vlined]{algorithm2e}
\makeatletter

\newcommand{\Rmnum}[1]{\expandafter\@slowromancap\romannumeral #1@}
\makeatother
\begin{document}
\title{\large Stochastic approach to study control strategies of Covid-19 pandemic in India }
\author{Athokpam Langlen Chanu and R.K. Brojen Singh}
\email{brojen@jnu.ac.in (Corresponding author)}
\affiliation{School of Computational \& Integrative Sciences, Jawaharlal Nehru University, New Delhi-110067, India.}
\begin{abstract}
{\noindent}India is one of the worst affected countries by the Covid-19 pandemic at present. We studied publicly available data of the Covid-19 patients in India and analyzed possible impacts of quarantine and social distancing within the stochastic framework of the SEQIR model to illustrate the controlling strategy of the pandemic. Our simulation results clearly show that proper quarantine and social distancing should be maintained from an early time just at the start of the pandemic and should be continued till its end to effectively control the pandemic. This calls for a more socially disciplined lifestyle in this perspective in future. The demographic stochasticity, which is quite visible in the system dynamics, has a critical role in regulating and controlling the pandemic.\\

\noindent{{\it \textbf{Keywords:}} Stochastic modelling; Covid-19; Stochastic Simulation Algorithm; Quarantine; Social distancing.}
\end{abstract}
%PACS number(s): 71.23, 72.15.R, 73.50
\maketitle
\vskip 0.5cm
\noindent\textbf{\large Introduction}\\
{\noindent}The peculiar nature of Covid-19 disease is the special form of pneumonia it causes and its fast spreading in the entire population. This dreaded Covid-19 disease is caused by the SARS-CoV-2 virus, and the disease outbreak was announced as a pandemic of special health attention by the World Health Organization (WHO) on 11 March, 2020. As of June 2, 2020 (10:06am CEST), there is a total Covid-19 confirmed case of 6,140,934 with 373,548 deaths all over the world \cite{who}.  India is one of the countries which is now seriously affected by this pandemic. As of June 2, 2020 (7:45pm IST), there are a total of 2,01,341 confirmed cases with 99,135 active cases and 5,632 deaths in India \cite{india}. The present situation of the outbreak is alarming since there is no vaccine/drug so far to cure this disease. The Indian Government, both at the central and state levels, has taken up special measures such as quarantine, social distancing and lockdown to prevent/intervene this pandemic across the country.\\

{\noindent}Mathematical epidemic models play important roles to study and predict disease dynamics as well as to implement necessary intervention strategies to control disease outbreak. Classic compartmental models such as the SI, SIS, SIR \cite{kk}, SEIR and their derived/extended models \cite{hh} have long been successfully used to study various disease transmission dynamics for different viruses such as H1N1 virus \cite{h1n1}, Ebola virus \cite{ebola}, SARS-CoV \cite{sars}, MERS-CoV \cite{mers}, etc. With special reference to the ongoing Covid-19 pandemic, there have been attempts using statistical methods, deterministic compartmental modeling, large scale simulation to study the Covid-19 disease dynamics in order to assist in mitigating the disease outbreak in several countries across the world. There have also been various studies in India regarding this Covid-19 outbreak in the country using various mathematical models which are mostly deterministic models \cite{1,2,3,4,5,6,7,9,10,11,12,13}. On the other hand, to capture the qualitative as well as quantitative real dynamical situations to intervene disease outbreak, stochastic approach needs to be employed. With the increasing capacity of modern computers, stochastic methods are gaining popularity because of being a powerful means to study and predict any complex dynamical system inherent with various environmental fluctuations/noise. In this work, we study the Covid-19 disease dynamics in India and some of its states using the classic SEQIR model from stochastic dynamics approach. Using stochastic numerical simulation, we illustrate in a very simplified manner the important impacts of social distancing and quarantine on the disease spreading mechanism specially in India and few states. We also highlight the importance of demographic stochasticity in the disease spreading dynamics.\\

\begin{figure*}[h]\includegraphics[height=7cm,width=13cm,angle=0]{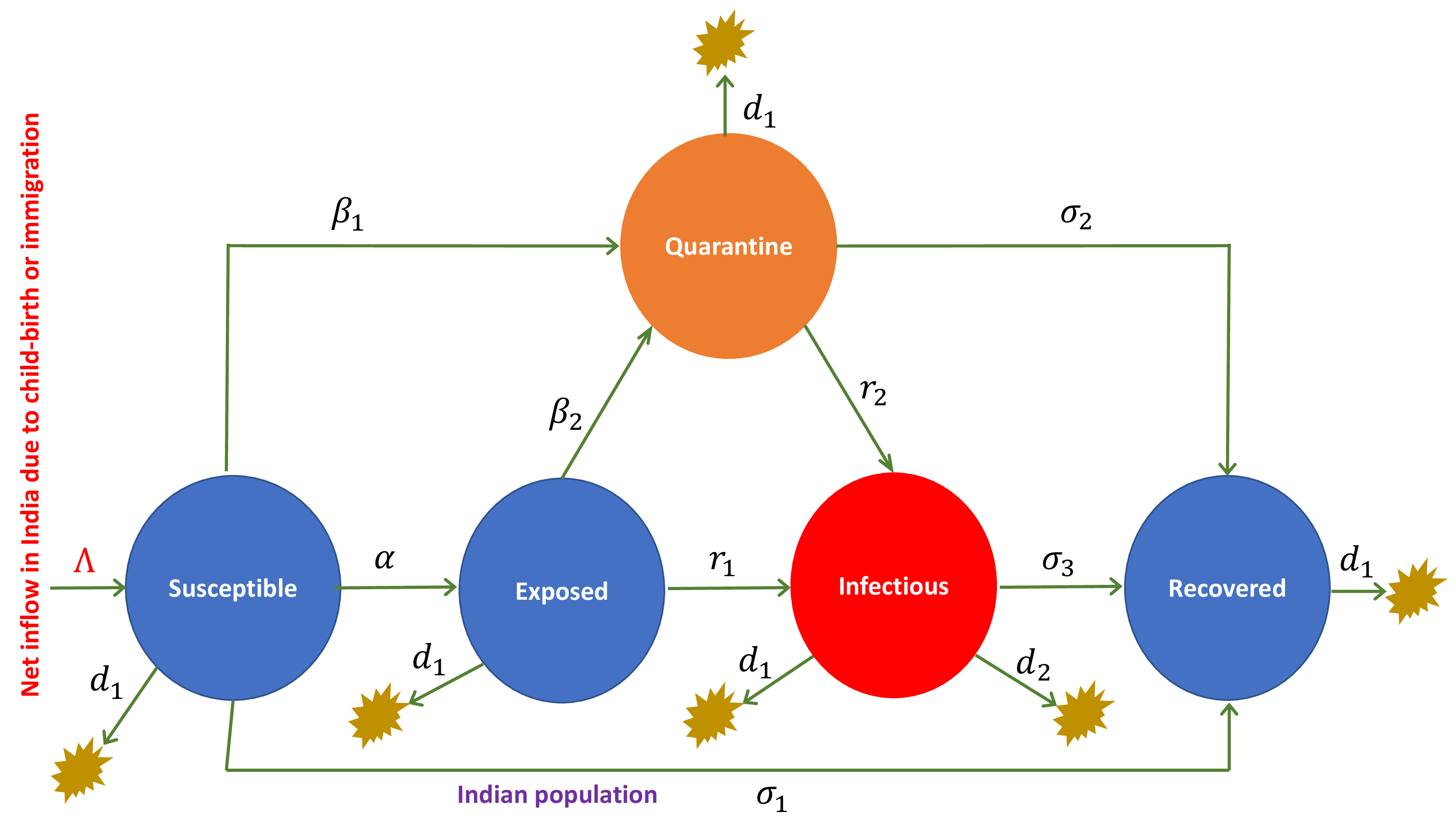}
\caption{The schematic diagram of SEQIR model.}
\label{fig1}
\end{figure*}

\vskip 0.3cm
\noindent\textbf{\large Methods}\\
{\noindent}The classical SEIR (Susceptible, Exposed, Infected and Recovered) model \cite{hh} was extended with another sensitive compartment called quarantine and the model is known as SEQIR model \cite{9} as shown in Fig. \ref{fig1}. In this model, the total population is sub-divided into five sub-populations (compartments) such as susceptible $(S)$; exposed and infected but not detected by testing $(E)$; self and/or institutional quarantined $(Q)$; confirmed/reported/hospitalized infected population $(I)$ and disease-recovered as well as population living in secured zone not affected by the Covid-19 outbreak $(R)$. We assume uniform mixing or homogeneity in the large population and also assume the total population at any instant of time is, $N= S(t) + E(t) + Q(t) + I(t) + R(t)$. Further, the change in population is in discrete integer amounts and the process of changing is a stochastic Markov process. Hence, the time evolution of each variable in the model should be considered in discrete and stochastic fashion \cite{gillespie}. In stochastic formalism of the model system [Fig. 1 and set of coupled differential equations (1)-(5)], the population state vector at any instant of time can be represented by, ${\bf X}=[S,E,Q,I,R]^{-1}$, which undergo $M=15$ reaction channels defined by, $\displaystyle\sum_{i=1}^{5}a_iX_i\stackrel{k_j}{\rightarrow}\sum_{i=1}^{5}b_iX_i$, where, $\{X_i\}=[S,E,Q,I,R];i=1,2...,5$. Here, $\{a\}$ and $\{b\}$ are the sets of reactant and product molecules respectively, and $\{k\}$ is the set of the classical rate constants. Further, the classical rate constants, ${k_j}$ can be related to the stochastic rate constants, ${c_j}$ by $c_j=k_jV^{1-\nu}$, where $\nu$ is the stoichiometric ratio and $V$ is the system size \cite{gillespie,gillespie1}. This incorporates the idea of correlating fluctuations in the dynamics of the system \cite{gillespie,dt,gillespie1}. Now, the reaction channels can be translated as,
\begin{eqnarray}
\label{reaction}
&&S + E\stackrel{\alpha}{\rightarrow}2E;~~
S\stackrel{\beta_1}{\rightarrow} Q;~~
S\stackrel{\sigma_1}{\rightarrow}R;~~
E\stackrel{r_1}{\rightarrow}I;~~
E\stackrel{\beta_2}{\rightarrow}Q;~~
Q\stackrel{\sigma_2}{\rightarrow}R;~~
Q\stackrel{r_2}{\rightarrow}I;~~
I\stackrel{\sigma_3}{\rightarrow}R\nonumber\\
&&I\stackrel{d_2}{\rightarrow}\phi;~~
\phi\stackrel{\Lambda}{\rightarrow}S;~~
S\stackrel{d_1}{\rightarrow}\phi;~~
E\stackrel{d_1}{\rightarrow}\phi;~~
Q\stackrel{d_1}{\rightarrow}\phi;~~
I\stackrel{d_1}{\rightarrow}\phi;~~
R\stackrel{d_1}{\rightarrow}\phi
\end{eqnarray}
where, $\alpha$ is the rate of disease transmission from the Susceptible to the Exposed and infected but not detected by testing class; $\beta_1$ is the rate of transition from the Susceptible to the Quarantined class; $\sigma_1$ is the rate of transition from the Susceptible to the secured zone class; $r_1$ is the rate of transition from the Exposed and infected but not detected by testing to the hospitalized Infected class; $\beta_2$ is the rate of transition from the Exposed and infected but not detected by testing to the Quarantine class; $\sigma_2$ is the rate of transition from Exposed and infected but not detected by testing to the secured zone class; $r_2$ is the rate of transition from Quarantine to hospitalized Infected class; $\sigma_3$ denotes the rate of transition from Quarantine to the secured zone class; $d_2$ indicates the rate of Covid-19 induced death; $\Lambda$ is the rate at which new individuals enter the Indian population due to a new child-birth or immigration in the country; and $d_1$ denotes the natural death rate.\\

{\noindent}If the system is subjected to a certain temperature $T$, the trajectory of any variable in \textbf{X} suffers a set of random molecular events given by the set of reactions (\ref{reaction}) and follows Brownian motion \cite{gillespie,dt,gillespie1}. Further, every time any one of the reaction sets is encountered, creation and annihilation of the molecular species will take place and hence, the state vector \textbf{X} will get changed as a function of time. Consider the state change from state \textbf{X} to another state \textbf{X$^\prime$} during the time interval $[t,t+\Delta t]$, then the time evolution of the configurational probability of state change $P(X;t)$ is given by the following Master equation constructed from the detailed balance equation \cite{gardiner,mc,van},
\begin{eqnarray}
\label{Master}
\frac{\partial P(S,E,Q,I,R;t)}{\partial t}&=&\alpha (S+1)(E-1) \ P(S+1,E-1,Q,I,R;t)+ \beta_1 (S+1)  \ P(S+1,E,Q-1,I,R;t)\nonumber \\&&+\sigma_1 (S+1) \ P(S+1,E,Q,I,R-1;t)+ r_1 (E+1) \ P(S,E+1,Q,I-1,R;t)\nonumber \\&&+ \beta_2(E+1) \ P(S,E+1,Q-1,I,R;t)+  \sigma_2 (Q+1)  \ P(S,E,Q+1,I,R-1;t) \nonumber \\&&+ r_2 (Q+1)  \ P(S,E,Q+1,I-1,R;t)+  \sigma_3 (I+1) \ P(S,E,Q,I+1,R-1;t)\nonumber \\&&+d_2 (I+1) \ P(S,E,Q,I+1,R;t)+  \Lambda  \ P(S-1,E,Q,I,R;t)\nonumber \\&&+ d_1 (S+1) \ P(S+1,E,Q,I,R;t)+ d_1 (E+1) \ P(S,E+1,Q,I,R;t)  \nonumber \\&&+ d_1 (Q+1) \ P(S,E,Q+1,I,R;t) + d_1 (I+1) \ P(S,E,Q,I+1,R;t) \nonumber \\&&+ d_1 (R+1) \ P(S,E,Q,I,R+1;t)-\left[\alpha SE + \beta_1 S+\sigma_1 S+r_1E\right.\nonumber \\&&+ \beta_2 E+\sigma_2Q+r_2 Q + \sigma_3 I+d_2I+ \Lambda + d_1 S+d_1 E  +d_1 Q \nonumber \\ &&\left.+ d_1 I+d_1 R\right] \ P(S,E,Q,I,R;t)
\end{eqnarray}
The Master equation (\ref{Master}) for complex multivariate systems is generally difficult to solve except for simple ones. However, the Master equation of any complex system can be solved numerically using the stochastic simulation algorithm (SSA) which is discussed briefly here. SSA is generally known as Doob-Gillespie algorithm, formulated by Gillespie \cite{gillespie,dt} based on the theoretical foundations developed by Doob JL \cite{Doob1,Doob2} and originally proposed by Kendall \cite{Kendall}. It is a Monte-Carlo type of algorithm, which is a non-spatial individual based analog of the Master equation incorporating all possible interactions in the system \cite{gillespie}. The SSA is built on two independent processes which are random viz. firing reaction and reaction time. These independent processes are realized by the generation of two uniform random numbers $r_1$ and $r_2$ which are statistically independent. The reaction time is computed using $\tau=-\frac{1}{a_0}ln(r_1)$, where, $a_0=\displaystyle\sum_{i}a_i$, where $a_i$ is the $i^{th}$ propensity function given by $a_i=h_ic_i$, where, $h_i$ is the number of possible molecular combinations of $i^{th}$ reaction. The $j^{th}$ reaction will fire when it satisfies, $\displaystyle\sum_{i=1}^{j}a_i\le a_0r_2\displaystyle <\sum^{j+1}_{i=1}a_i.$ \\

\vskip 0.3cm
\noindent\textbf{\large Results and Discussion}\\
{\noindent}We present below the simulation results of the model proposed, discuss the analysis and provide possible prediction of the Covid-19 pandemic in India and few states which are affected seriously. In population dynamics, one can express $V$ as , $V=\frac{N}{N/V}=\frac{N}{D}$; where, $N$ is the total population in the geographical area, and $D$ is the population density. Taking $N$ as constant, we can correlate the change in $V$ as change in the $D$ by, $V\propto\frac{1}{D}$.\\

\begin{figure*}[t]\includegraphics[height=11cm,width=15cm]{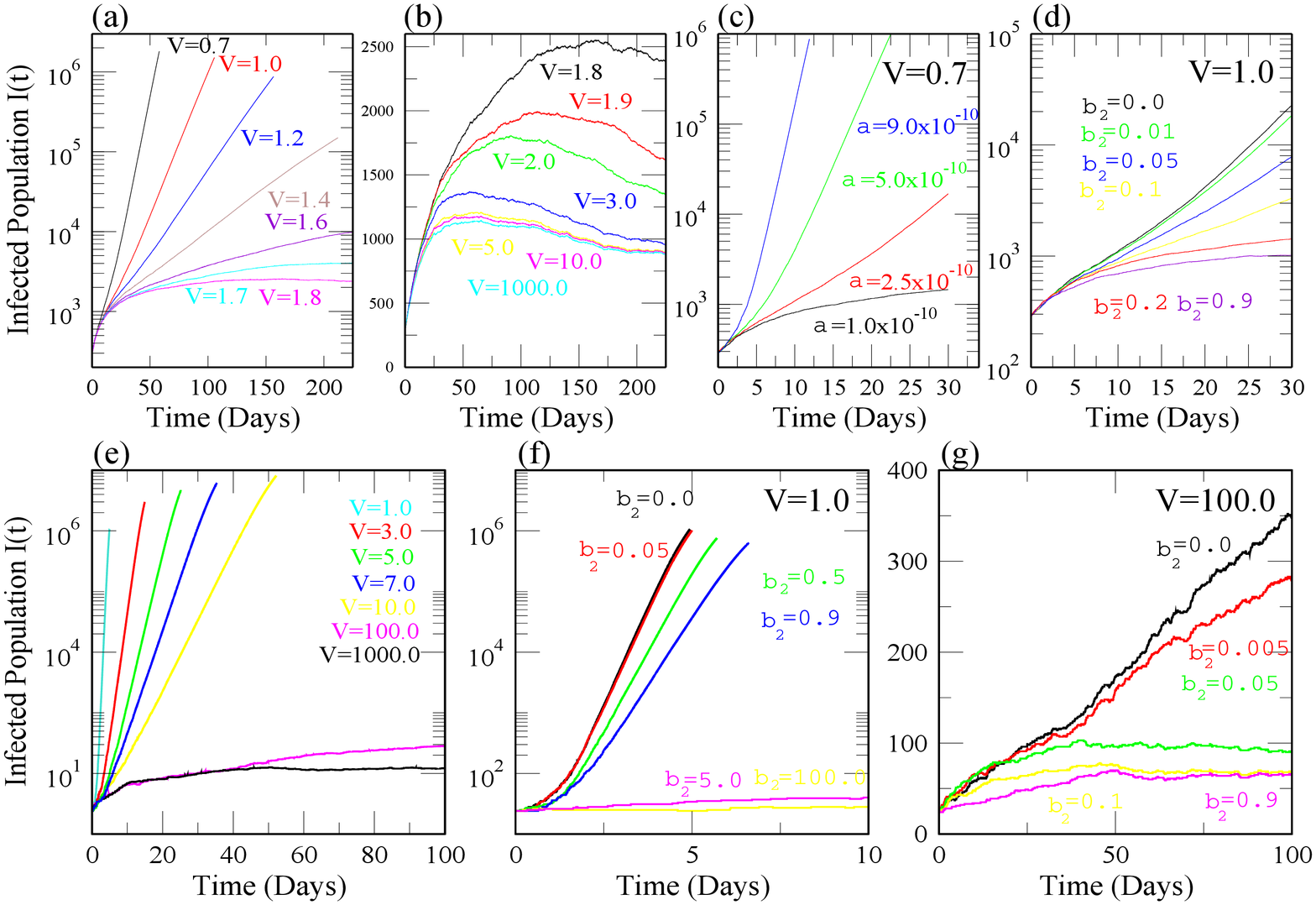}
\caption{\small \sl \textit{The upper panels (a),(b),(c) and (d) represent the Simulation Results of Infected Population $I(t)$ vs time $t$ (in days) for India using Stochastic Simulation Algorithm. (a) and (b) show $I(t)$ vs $t$ for India for different values of $V$. (c) shows $I(t)$ vs $t$ for India for different values of transmission rate $\alpha$ at $V=0.7$. (d) shows $I(t)$ vs $t$ for India for different values of quarantine rate $\beta_2$ at $V=1.0$. The lower panels (e),(f) and (g) represent the Simulation Results of Population $I(t)$ vs time $t$ (in days) for Uttar Pradesh using Stochastic Simulation Algorithm. (e) shows $I(t)$ vs $t$ for Uttar Pradesh for different values of $V$. (f) and (g) show $I(t)$ vs $t$ for Uttar Pradesh for different values of quarantine rates $\beta_2$ at two fixed volumes $V=1.0$ and $V=100.0$ respectively. }}  \label{fig:fig2}
\end{figure*}

\noindent\textbf{Scenario 1 (India):} We numerically simulate the time-evolution of the infected population $I(t)$ in India using SSA under different conditions. The rate constants values and initial values for the simulation are taken from \cite{9} by verifying to the present data from March 21, 2020 to May 31, 2020 \cite{india}. The population of Covid-19 infected people in India, $I(t)$, is greatly effected by population density $D$ [Fig. \ref{fig:fig2} (a), (b)], and is not homogeneously distributed over India. From the simulation results, dynamics of $I$ for various values of $V$ (or $D$), it is observed that at values of $V<1.7~[D>\frac{N}{1.7}]$, $I(t)$ is increased exponentially indicating monotonic increase in infection if the density of population $D$ increases (Malthusian law \cite{Malthus} in the increase in infected population). The curves start flattening (Gompertz-Winsor nature \cite{Winsor}) around $V\geq 1.7~[D\leq\frac{N}{1.7}]$, which is the signature of endemic [Fig. \ref{fig:fig2} (b)], and the fluctuations in the dynamics due to $D$ playing important role in intervening the disease spreading. Hence, from these results it can be predicted that the controlling population density could lead to the endemic of Covid-19 pandemic in India. In other words, the strategy to decrease in $D$ could be isolation of susceptible and exposed from infectious people either by quarantine them (at home or isolated place) or lockdown of socially interacting places (academic institutions, offices, festivals etc).\\

{\noindent}In Fig. \ref{fig:fig2} (c), the sensitivity of the model is studied with respect to the parameter $\alpha$, which is the disease transmission rate from $S$ to $E$ population Fig. \ref{fig1}, at a particular value $V=0.7$. As the value of $\alpha$ increases, the population $I(t)$ increases sharply. This corresponds to the fact that for a certain demographic region, as the transmission rate increases, due to homogeneous mixing of population, more $S$ population gets exposed and then infected with the disease.  Further, we studied the impact of the quarantine rate, $\beta_2$ on $I(t)$, where, the quarantine rate is given by, $\beta_2 \sim \frac{1}{delay \ in \ quarantine}$. In Fig. \ref{fig:fig2} (d), we see that as the quarantine rate increases or delay in quarantine decreases, $I(t)$ also starts decreasing sharply. When there is no quarantine or $\beta_2=0.0$, the $I(t)\sim O(10^4)$. However, when  $\beta_2=0.2$ (quarantine in $5$ days), $I(t)$ starts flattening around $\sim O(10^3)$. This result illustrates the important effect of quarantine on $I(t)$ indicating quarantine of population $E$ needs to be done as quickly as possible and for longer time also for intervening the disease spreading. We know that flattening the curve can prevent the burden on hospitals and health care facilities which in turn will keep the pandemic under control.\\

\noindent We again perform numerical simulation of the evolution of the infected population $I(t)$ with respect to time $t$ in five Indian states namely, Uttar Pradesh, Delhi, Kerala, Maharashtra and West Bengal. These states are comparatively highly affected by Covid-19 pandemic and densely populated states among other Indian states. The rate constant values and initial values for the simulation of all five states are taken from \cite{9} after verification with the present case. \\

\noindent\textit{Scenario 2 (Uttar Pradesh):} It is observed in Fig. \ref{fig:fig2} (e), that at around $t=20 \ days$, $I(t)\sim O(10^5) \ (V=5~[D=\frac{N}{5}]), I(t)\sim O(10^4)\ (V=7~[D=\frac{N}{7}]), I(t)\sim O(10^3) \ (V=10,~[D=\frac{N}{10}])$. When $V\approx 100~[D\approx\frac{N}{100}]$, the $I(t)$ curve starts flattening indicating the intervention of the pandemic. Hence, the demographic stochasticity measured by $1/\sqrt{V}\propto\sqrt{D}$ can control the time evolution of $I(t)$, and consequently intervene disease spreading. Again, in Fig. \ref{fig:fig2} (f) and (g), we studied the effect of quarantine on $I(t)$ evolution for two values, $V=1.0~[D=N]$ and $V=100~[D=\frac{N}{100}]$. In Figure \ref{fig:fig2} (f), at $V=1~[D=N]$, quick quarantining $E$ population (in hours) is needed to prevent the disease outbreak and fast transmission of the disease. From Fig. \ref{fig:fig2} (g) for  $V=100.0$, it can be seen that if the exposed $E$ population is quarantined in ten days, the disease outbreak is relatively controlled.\\

\begin{figure*}[t]\includegraphics[height=18cm,width=18cm]{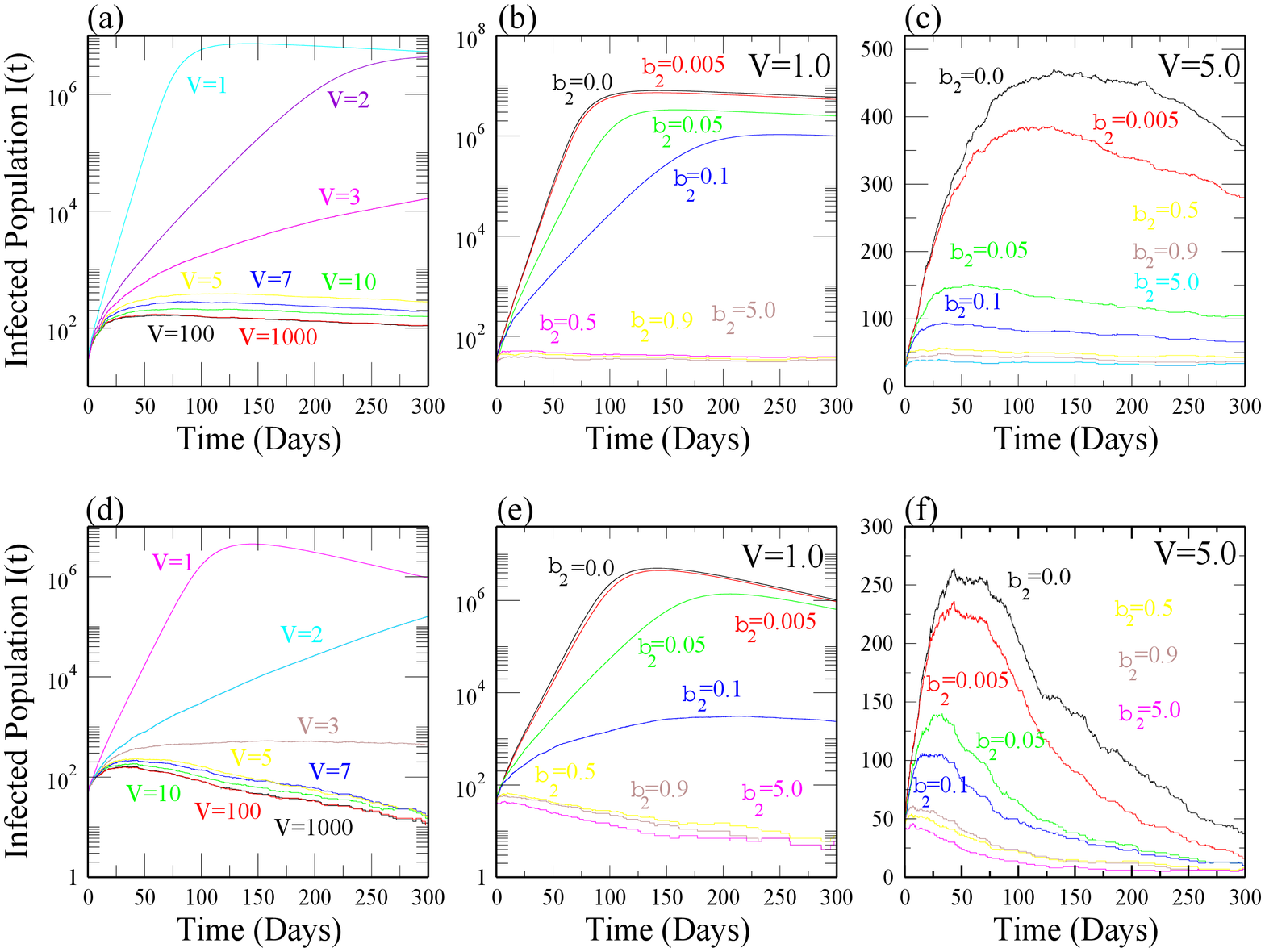}
\caption{\small \sl \textit{(a), (b) and (c) show simulation results of Infected Population $I(t)$ vs time $t$ in days for Delhi using Stochastic Simulation Algorithm. (a) shows $I(t)$ vs $t$ for different values of $V$. (b) and (c) shows $I(t)$ vs $t$ for different values of $\beta_2$ at two different volumes $V=1.0$ and $V=5.0$ respectively. Again, (d), (e) and (f) show simulation results of Infected Population $I(t)$ vs time $t$ in days for Kerala using Stochastic Simulation Algorithm. (d) shows $I(t)$ vs $t$ for different values of $V$. (e) and (f) shows $I(t)$ vs $t$ for different values of $\beta_2$ at two different volumes $V=1.0$ and $V=5.0$ respectively.   }} 
\label{fig:delhiandkerala}
\end{figure*}

\noindent\textit{Scenario 3 (Delhi):} From the simulation results of the dynamics of $I(t)$ for different values of $V$, it is seen that the magnitude of $I(t)$ varies drastically with the values of $V$. When $V=1~[D=N]$, $I(t)\sim O(10^6)$, but when $V\approx 5~[D\approx\frac{N}{5}], I(t)\sim O(10^2)$. We also observe that the flattening of the curve is achieved at earlier times of the disease outbreak if the values of $V$ is increased. This could be due to the fact that the number density $n$ of the population decreases when $V$ is increased as the total population $N$ is fixed. The decrease in $n$ may be considered due to policy like social distancing, isolation etc. Our simulation study also clearly shows that social distancing plays a crucial role in early times of the epidemic for proper intervention. Again, in Fig. \ref{fig:delhiandkerala} (b) and (c), the time evolution of $I(t)$ is studied at different quarantine rates at two different values $V=1$ and $V=5~[D=\frac{N}{5}]$. For the same parameter values, at $V=1~[D=N]$, $I(t)$ shows an exponential increase whereas when it is at $V=5~[D=\frac{N}{5}]$, $I(t)$ already starts flattening and decreasing (also see \ref{fig:delhiandkerala}(a)). In Fig. \ref{fig:delhiandkerala} (b) and (c), the magnitudes of $I(t)$ at both values of $V$ are seen to decrease as the quarantine rate $\beta_2$ increases. If the quarantine process of $E$ population is carried out as early as in two days, the disease spreading is greatly controlled (roughly 50 infected population in our simulation of the model in consideration). This highlights the importance of quarantine as early as possible in densely populated places during an epidemic. Hence, the simulation results based on the data indicates that the disease spreading in Delhi is still not controlled properly and needed to take up serious precautions in the state, such as, proper quarantine, social distancing, lockdown etc.\\

\noindent\textit{Scenario 4 (Kerala):} The results of Kerala are quite different from others [Fig. \ref{fig:delhiandkerala} (d), (e), (f)]. The dynamics of $I(t)$ for different  values of $V$ show that the population $I(t)$ decreases with increased in $V$. When $V=1~[D=N]$, $I(t)\sim O(10^6)$, whereas, for $V\approx 5~[D\approx\frac{N}{5}], I(t)\sim O(10^2)$. We observed an overall decline in $I(t)$ for all $V\ge 1~[D=N]$ leading to proper control of pandemic by monitoring infected population density $D$ as discussed in the above cases. We also found the effect of decreasing the number density $n$ of the population (may be due to policy, such as, social distancing as mentioned earlier) at early times of the epidemic could be a strategy for intervening the pandemic. Again, in Fig. \ref{fig:delhiandkerala} (e) and (f), we studied the impact of quarantine ($\beta_2$) on the time evolution of $I(t)$ for two different values $V=1~[D=N]$ and $V=5~[D=\frac{N}{5}]$. In Fig. \ref{fig:delhiandkerala} (e) and (f), the magnitudes of $I(t)$ at both values of $V$ decrease as the quarantine rate $\beta_2$ increases. In Fig. \ref{fig:delhiandkerala} (e), if the quarantine process of $E$ population is carried out in ten days, then the $I(t)$ curve becomes flatten, and, if the quarantine process is introduced in two days, then the $I(t)$ curve shows a decreasing trend. Further, in Fig. \ref{fig:delhiandkerala} (f), if the quarantine process of exposed $E$ population is carried out in two days, then the disease spreading is immensely controlled. This again highlights the importance of quarantine as early as possible during a disease outbreak. Hence, disease spreading in Kerala is quite controlled as compared to other Indian states.\\

\begin{figure*}[t]\includegraphics[height=18cm,width=18cm]{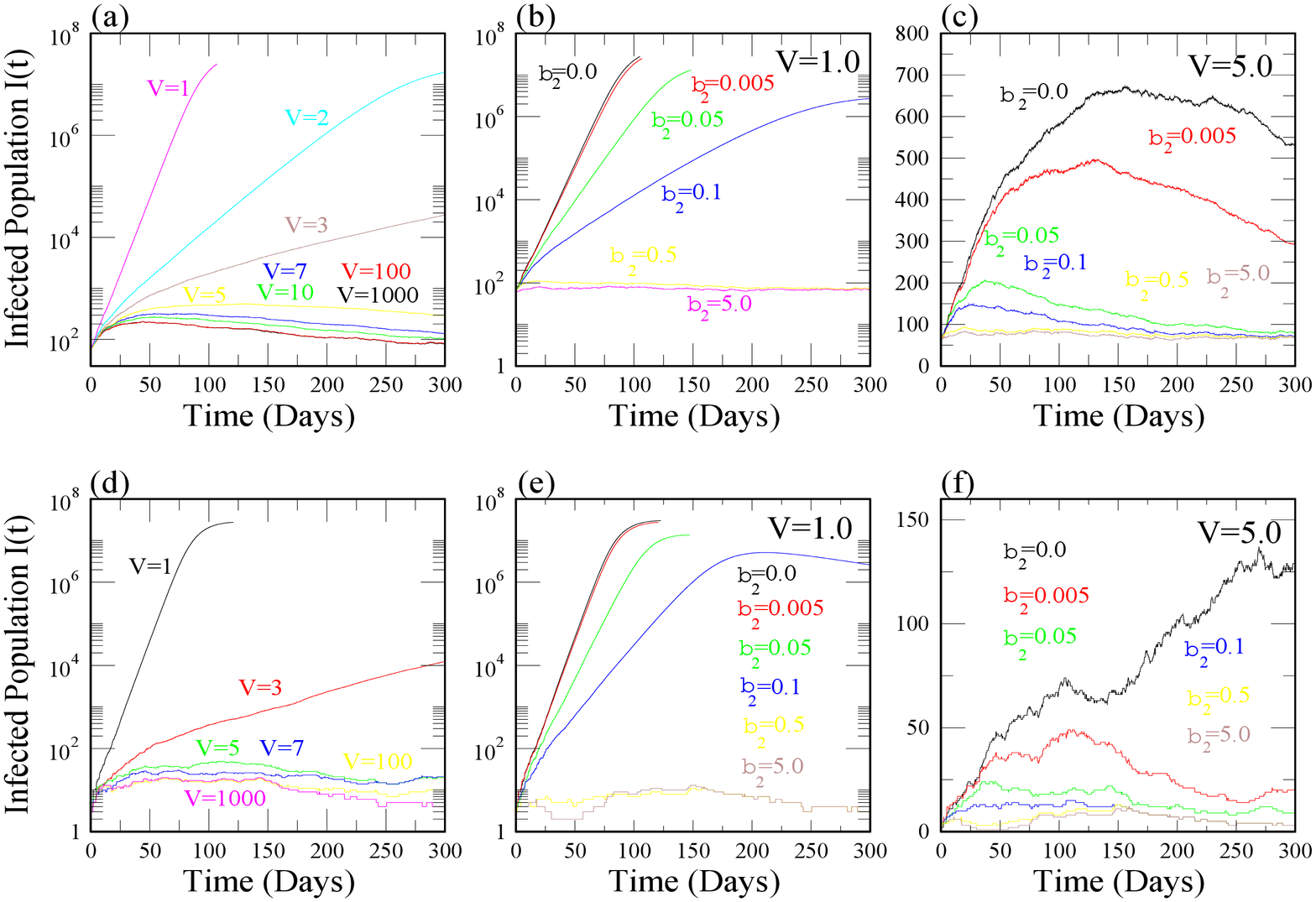}
\caption{\small \sl \textit{(a), (b) and (c) show simulation results of Infected Population $I(t)$ vs time $t$ in days for Maharashtra using Stochastic Simulation Algorithm. (a) shows $I(t)$ vs $t$ for different values of $V$. (b) and (c) shows $I(t)$ vs $t$ for different values of $\beta_2$ at two different volumes $V=1.0$ and $V=5.0$ respectively. Again, (d), (e) and (f) show simulation results of Infected Population $I(t)$ vs time $t$ in days for West Bengal using Stochastic Simulation Algorithm. (d) shows $I(t)$ vs $t$ for different values of $V$. (e) and (f) shows $I(t)$ vs $t$ for different values of $\beta_2$ at two different volumes $V=1.0$ and $V=5.0$ respectively.   }}
\label{fig:mahaandwb}
\end{figure*}

\noindent\textit{Scenario 5 (Maharashtra):} Now, Fig. \ref{fig:mahaandwb} (a) shows the dynamics of $I(t)$ for different values of $V$, and found that $I(t)$to is decreased with $V$. When $V=1~[D=N]$, peak $I(t)\sim O(10^7)$ indicating Malthusian character. However, when $V\approx 5~[D\approx\frac{N}{5}], peak~ I(t)\sim O(10^3)$ showing flattening of the curve, such that for $V\ge 5~[D\le\frac{N}{5}] $, the disease spreading is quite controlled. We, further, observed the effect of increasing the value of $V$ or decreasing the number density $n$ of the population in controlling the number of infected populations $I(t)$. Again, in Fig. \ref{fig:mahaandwb} (b) and (c), the time evolution of $I(t)$ is again studied at different quarantine rates for two fixed values $V=1~[D=N]$ and $V=5~[D=\frac{N}{5}]$. The results showed that for small values of $\beta_2$, the dynamics of $I(t)$ follow Malthusian law, whereas, for significantly large values of $\beta_2$, the curves just start flattening indicating pandemic controlled behavior. As obtained before, if the quarantine process of exposed $E$ population is carried out as early as possible during a disease outbreak, then the number of $I(t)$ can be systematically controlled. But still the condition of Maharastra state is alarming as compared to other state, and proper strategy needed to be taken up.\\

\noindent\textit{Scenario 6 (West Bengal):} In the case of simulation results of West Bengal based data [Fig. \ref{fig:mahaandwb} (d)], it is found that when $V=1~[D=N]$, peak $I(t)\sim O(10^8)$. However, for $V\approx 5~[D\approx\frac{N}{5}]$, we observed that peak $I(t)\sim O(10^2)$ indicating flattening of the curve, and for $V\rangle 5~[D\langle\frac{N}{5}]$ the $I(t)$ curves show decreasing nature which is the signature of controlling of disease spreading. Further, in Fig. \ref{fig:mahaandwb} (e) and (f), we studied the effect of quarantine on $I(t)$ evolution for two fixed values $V=1~[D=N]$ and $V=5~[D=\frac{N}{5}]$. From the results we observed that if proper quarantine of the exposed population (small values of $\beta_2\langle 0.5$) increase in population $I(t)$ is quite large and curve flattening takes long time (100-200 days). But proper quarantine of population $E$ is done (large values of $\beta_2\ge 0.5$), the dynamics of $I(t)$ became decreased, and pandemic can be controlled. The data based simulation indicates that the scenario of West Bengal is also still alarming as compared to other Indian states.\\

\begin{figure*}[t]\includegraphics[height=9cm,width=18cm]{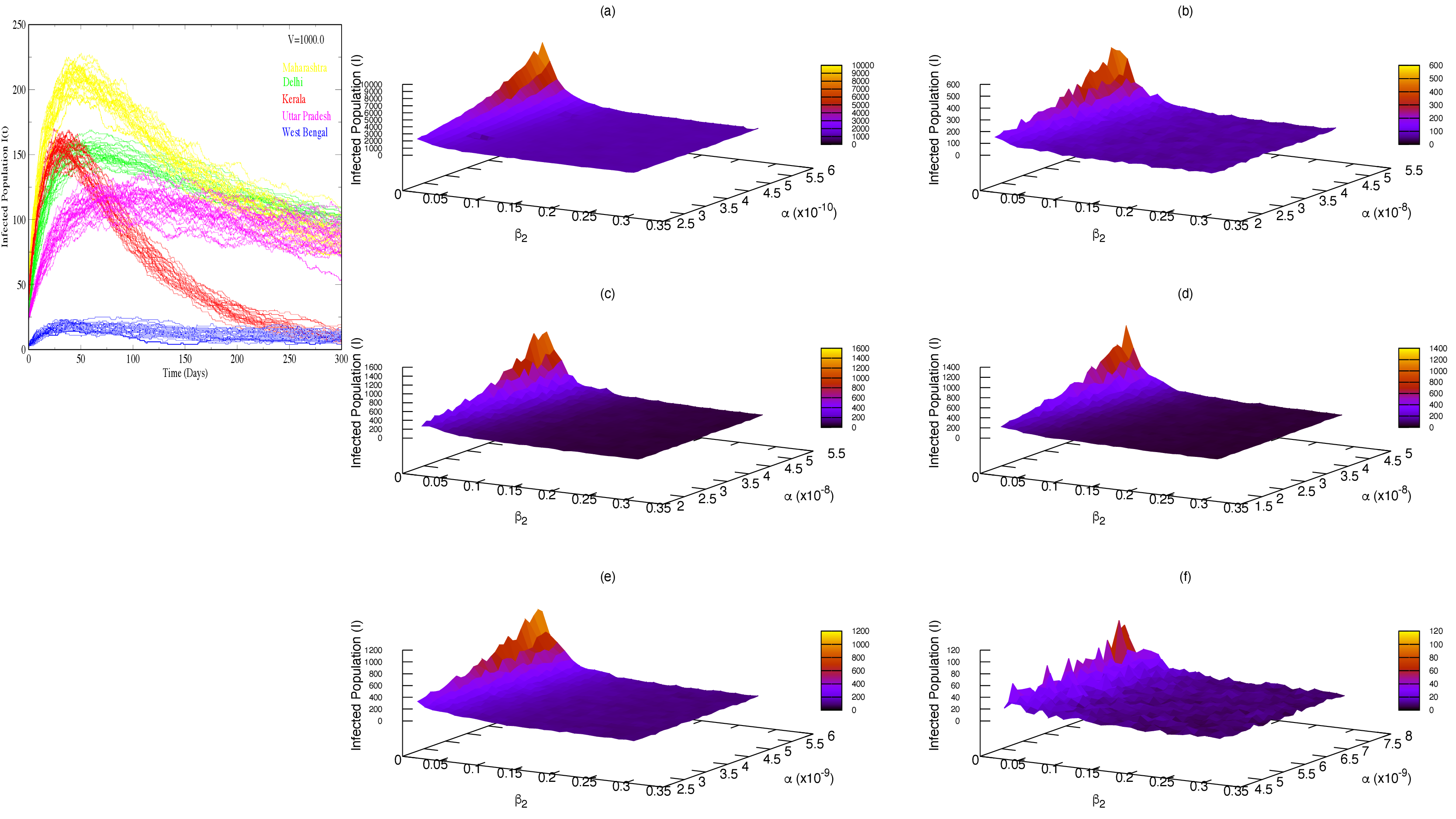}
\caption{\small \sl \textit{Simulation results of $I(t)$ vs time $t$ (in days) for five Indian states Maharashtra, Delhi, Kerala, Uttar Pradesh and West Bengal at a fixed volume $V=1000.0$. Variation of Infected population w.r.t transmission rate $\alpha$ and quarantine rate $\beta_2$: Result of Heat Map for (a) India, (b) Uttar Pradesh, (c) Delhi, (d) Kerala, (e) Maharashtra, and (f) West Bengal.}}  \label{fig:fivestates}
\end{figure*}

\noindent\textit{Disease spreading pattern:} We, then, studied the how Covid-19 spread in the parameter space $(\alpha,\beta_2,I)$ as shown in Fig. \ref{fig:fivestates}. The dynamics of $I(t)$ for all the five Indian states at one particular value of $V=1000$ for thirty realization each show that the trajectories show peaks around $50$ days [Fig. \ref{fig:fivestates} left panel] and start declining. The dynamics show that Covid-19 spreading in Kerala state is quite controlled as compared to other states. Then we calculated $I(t)$ after simulating hundred days in the parameter space $(\alpha,\beta_2,I)$ for India and her five states [Fig. \ref{fig:fivestates} (a)-(f)]. From the plots, we observed that the population $I$ is relatively large for a large value of $\alpha$ and small value of $\beta_2$. As the value of $\beta_2$ increases, the $I$ population drops, and stochastic fluctuations can be seen in these heat maps. Hence, in order to control the Covid-19 pandemic, the parameters $\alpha$ and $\beta_2$ need to be optimized.\\  
\vskip 0.3cm

\noindent\textbf{\large Conclusion}\\
{\noindent}We have studied the stochastic SEQIR model in the context of Covid-19 disease dynamics in India and its five comparatively worse affected states using stochastic methods. Our numerical simulation results show that policies like social distancing and quarantine have important roles in controlling the disease outbreak and we propose to optimize these two parameters to effectively intervene in the disease transmission. An important consequence of employing a stochastic method is the importance of demographic fluctuations, which are quite visible in the simulation results, to affect the disease dynamics and even in intervening the disease spread. This demographic stochasticity is generally neglected in its deterministic counterpart which is quite important in regulating any system dynamics. Hence, our stochastic simulation method could capture the demographic stochasticity which is non-negligible. We would like to mention that we do not intend to give quantitative predictions here. One limitation of the model under consideration is that by construction, populations from $S$ and $E$ compartments make transitions to the $Q$ compartment where they are assumed to interact homogeneously. This may give rise to a more infected population and we do not see the trends of $I(t)$ converging near zero over relatively less $300$ days in our simulation. This also points out that policies such as social distancing and quarantine of the exposed population are not sufficient enough to end the COVID-19 disease outbreak. Other policies like complete lockdown and more testing of susceptible populations should be considered and must be incorporated systematically in mathematical models. 
\vskip 0.3cm
\noindent\textbf{Authors Contribution}\\ 
The conceptualisation of the present work is done by RKBS and ALC. Both authors carried out the numerical simulation as well as the preparation of associated figures. Both authors wrote, discussed and approved the final manuscript.
\vskip 0.3cm
\noindent\textbf{Competing financial interests}\\
\noindent The authors declare no competing financial interests.
\vskip 0.3cm
\noindent\textbf{Acknowledgments}\\
\noindent ALC is a DST-Inspire Fellow (IF180043) and acknowledges Department of Science and Technology (DST), Government of India for financial support under Inspire Fellowship scheme (order no:DST/INSPIRE Fellowship/[IF180043]). RKBS acknowledges DBT-COE, India, for providing financial support.

\vspace{0.5cm}
%\section*{References}

\begin{table}[h]
\caption{Rate constant values taken from \cite{9}}
\centering
\begin{tabular}{|c| c| c| c| c| c| c| }
\hline
\textbf{Rate constants ($day^{-1}$)} & \textbf{India} & \textbf{Maharashtra}  & \textbf{Kerala} & \textbf{Delhi} & \textbf{Uttar Pradesh} & \textbf{West Bengal}\\
[1ex]
\hline
$\alpha$ &   $0.00000000025$ &   $0.0000000025$ &   $0.0000000164$ &   $0.000000021$ &   $0.00000002$ &   $0.0000000047$ \\
[1ex]
\hline
$\beta_1$ &   $0.0000004$ &   $0.0000004$ &   $0.0000004$ &  $0.0000004$ &   $0.0000004$&  $0.0000004$ \\
[1ex]
\hline
$\sigma_1$&   $0.0005$ &    $0.0005$ &    $0.0005$ &    $0.0005$ &    $0.0005$ &    $0.0005$ \\
[1ex]
\hline
$r_1$ &   $0.01$ &    $0.04$  &    $0.05$  &   $0.05$  &    $0.02$  &    $0.05$  \\
[1ex]
\hline
$\beta_2$ &   $0.1$ &    $0.005$ &    $0.005$ &   $0.005$ &    $0.005$ &    $0.005$ \\
[1ex]
\hline
$\sigma_2$&   $0.05$ &   $0.1$ &   $0.1$ &   $0.1$ &   $0.1$ &   $0.1$ \\
[1ex]
\hline
$r_2$ &   $0.001$ &  $0.002$ &   $0.001$ &  $0.00093$ &   $0.0012$ &   $0.0005$ \\
[1ex]
\hline
$\sigma_3$ &   $0.006$ &    $0.005$ &    $0.012$ &    $0.0016$ &    $0.0038$ &    $0.0078$ \\
[1ex]
\hline
$d_2$ &   $0.00197$ &   $0.0032$  &    $0.00029$  &    $0.00067$ &    $0.0049$  &    $0.0024$  \\
[1ex]
\hline
$\Lambda$ &   $40000$ &   $3300$ &   $405$ &   $650$ &   $14200$ &   $3490$ \\
[1ex]
\hline
$d_1$ &$0.00002$ &   $0.000015$ &   $0.000018$ &   $0.00001$ &   $0.00002$ &     $0.000016$ \\
[1ex]
\hline
\end{tabular}
\label{table3}
\end{table}
\begin{table}[h]
\caption{Initial values taken from \cite{9}}
\centering
\begin{tabular}{|c| c| c| c| c| c| c| }
\hline
\textbf{Initial Values} & \textbf{India} & \textbf{Maharashtra}  & \textbf{Kerala} & \textbf{Delhi} & \textbf{Uttar Pradesh} & \textbf{West Bengal}\\
[1ex]
\hline
$S(0)$ &   $800000000$ &   $75000000$ &   $10000000$ &   $10000000$ &   $150000000$ &   $50000000$ \\
[1ex]
\hline
$E(0)$ &   $1500$ &   $225$ &   $200$ &  $150$ &   $120$&  $20$ \\
[1ex]
\hline
$Q(0)$&   $50000$ &    $800$ &    $1000$ &    $800$ &    $1500$ &    $200$ \\
[1ex]
\hline
$I(0)$ &   $284$ &    $58$  &    $40$  &   $27$  &    $24$  &    $3$  \\
[1ex]
\hline
$R(0)$ &   $400000000$ &    $30000000$ &    $10050000$ &   $500000$ &    $50000000$ &    $30000000$ \\
[1ex]
\hline
\end{tabular}
\label{table4}
\end{table}
\newpage


\end{document}